**І. О. Теплицький, С. О. Семеріков**

*Криворізький національний університет*

# МОДЕЛЮВАННЯ ЗА ДОПОМОГОЮ ВИПАДКОВИХ ЧИСЕЛ

Статтю присвячено методиці побудови та дослідження стохастичних моделей на основі методу Монте-Карло. Розглядається модель броунівського руху, побудова й опрацювання якої вводить у світ випадкових чисел і математичної статистики, сприяє формуванню уявлень про розподіли ймовірностей, зокрема ілюструє два поширених розподіли: рівномірний та нормальний.

**Ключові слова**: комп'ютерне моделювання, метод Монте-Карло, випадкові числа, рівномірний розподіл, електронні таблиці.

**Постановка проблеми.** Зазвичай перебіг багатьох процесів визначається строгими й чіткими закономірностями: значення вихідних параметрів однозначно залежать від значень відповідних величин на вході (їх початкових значень). Ці закономірності подаються математичним записом у вигляді точних формул. Явища, що описуються такими величинами, мають назву детермінованих (від латинського *determino* – визначати, обумовлювати), таку ж назву мають і відповідні моделі. Проте, окрім детермінованих процесів і явищ, існують і такі, що для них неможливо за допомогою точних формул врахувати різноманітні впливи випадкових факторів. Їхні характеристики за своєю природою можуть набувати лише випадкових значень. Такі величини називають випадковими або *стохастичними* (від грецького *stochasticos* – той, що вміє вгадувати, випадковий). Цю ж назву – «стохастичні» – мають і математичні моделі, що містять такі величини. Якщо в детермінованих явищах багаторазово відтворювати ті самі початкові умови, то обов'язково відтворюватимуться ті самі результати. У випадку стохастичних процесів результати кожного разу будуть новими.

**Аналіз останніх досліджень з вирішення загальної проблеми та виділення невирішених питань.** У роботах [2–7] розглянуті основні елементи педагогічної технології комп'ютерного математичного моделювання, систематично викладеної у навчальному посібнику [8]; наводяться численні приклади її застосування до побудови й дослідження детермінованих навчальних моделей у середовищі електронних таблиць.

**Метою статті** є розгляд методики побудови стохастичних моделей.

**Виклад основного матеріалу**

**1. Метод Монте-Карло.**

Існують різні підходи до моделювання систем, що містять стохастичні характеристики, але найбільш простим і поширеним є метод випадкової вибірки або метод Монте-Карло. Його назва походить від назви столиці князівства Монако, відомої в усьому світі своїми гральними домами, де чільне місце посідає рулетка. Якщо рулетка гарно збалансована, кулька може зупинитись у любому положенні, тому ймовірність одержання будь-якого числа однакова для всіх чисел на барабані. Це приклад так званого *рівномірного розподілу* випадкових величин. У реальних (природних, виробничих, суспільних) явищах спостерігаються розподіли нерівномірні. Вони характерні для коливань купівельного попиту, для величини врожаю в різні роки, для виробничих похибок та похибок вимірювань, для рівня перешкод при передаванні інформації тощо. Всіх їх вивчає окрема теорія – *математична статистика*.

Ідея методу Монте-Карло полягає в тім, що при побудові стохастичних моделей деякі параметри моделі визначають за допомогою випадкових чисел. Основна проблема тут зводиться до пошуку зручного й надійного джерела (генератора) таких чисел. За наявності комп'ютера користуються стандартним генератором *псевдовипадкових* чисел.

**2. Моделювання броунівського руху (найпростіша модель).**

Пригадаймо, що броунівським називають безладний рух дрібних частинок, завислих у рідині чи газі. Як було встановлено, причиною руху броунівської частинки є відсутність точної просторової компенсації ударів, що їх зазнає частинка з боку оточуючих її молекул внаслідок їх теплового руху. Ці некомпенсовані удари приводять частинку у невпорядкований рух: швидкість її весь час різко змінюється і за величиною, і за напрямком. Якщо фіксувати положення довільної частинки через невеликі однакові проміжки часу, то побудована в такий спосіб траєкторія виявляється надзвичайно складною й заплутаною ламаною лінією. На *рис. 1* показані фотографії траєкторій рухів трьох броунівських частинок радіусом 0,52 мкм у воді [1]. Точками відмічені положення частинок через кожні 30 с. Відстань між поділками сітки 3,4 мкм.

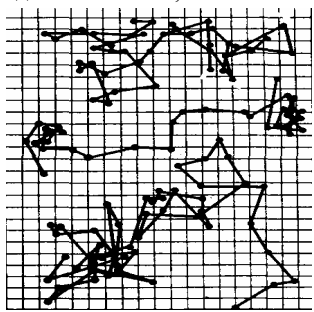

*Рис. 1*

**2.1. Комп'ютерна модель броунівського руху.** Значення проекцій переміщень $s_x$ і $s_y$ броунівської частинки будемо моделювати парами випадкових чисел, які в середовищі електронних таблиць в інтервалі [0; 1] продукує функція СЛЧИС (*рис. 2*). Оскільки всі напрямки руху однаково ймовірні то для того, щоб ці проекції могли набувати як додатних значень, так і від'ємних, випадкові числа мають змінюватись від –1 до +1. Такі числа даватиме функція 2*СЛЧИС( ) – 1 (*доведіть!*).

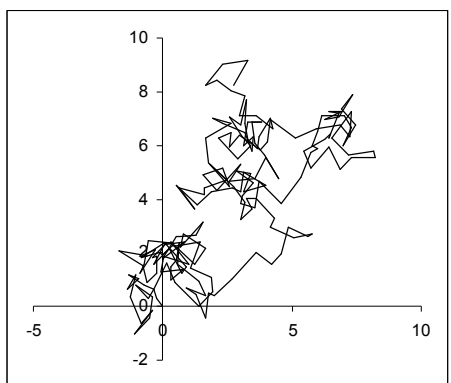

*Рис. 2*

Нову координату $x_{i+1}$ частинки будемо знаходити, додаючи до її попередньої координати $x_i$ відповідну проекцію переміщення $s_{xi}$: $x_{i+1} = x_i + s_{xi}$.

Крім того будемо обчислювати модулі проекцій переміщення $|s_{xi}|$, $|s_{yi}|$ і модуль вектора переміщення $|s_i| = \sqrt{s_{xi}^2 + s_{yi}^2}$.

**2.2. Обговорення алгоритму роботи з моделлю.**

1. Створимо таблицю за таким зразком

|   | A | B | C | D | E | F | G |
|---|---|---|---|---|---|---|---|
| **1** | $x$ | $y$ | $s_x$ | $s_y$ | $|s_x|$ | $|s_y|$ | $|s|$ |
| **2** |   |   |   |   |   |   |   |
| **...** | ... | ... | ... | ... | ... | ... | ... |

2. У першому рядку помістимо імена змінних: $x$, $y$ – координати частинки; $s_x$, $s_y$ – проекції переміщення $s$ на координатні осі; $|s_x|$, $|s_y|$ – модулі проекцій переміщення на ці осі; $|s|$ – модуль вектора переміщення. Для визначеності початкові координати частинки приймемо рівними нулю: $x_0 = 0$, $y_0 = 0$. Вміст комірок має бути наступним:

| Комірки | Формули / числа |
|---|---|
| A2 | =0 |
| A3 | =A2+(2*СЛЧИС()-1) |
| B2 | =0 |
| B3 | =B2+(2*СЛЧИС()-1) |
| C2 | пуста |
| C3 | =A3-A2 |
| D3 | =ABS(C3) |

3. Комірки E3, F3 і G3 заповнити самостійно.
4. Третій рядок копіюємо в наступні 100 рядків, тобто до рядка з номером 102 включно.
5. За даними стовпців A і B побудуємо траєкторію руху частинки, тобто графік залежності координати $y$ від координати $x$.

**2.3. Обчислювальний експеримент.**

|   | A | B | C | D | E | F | G |
|---|---|---|---|---|---|---|---|
| **1** | $x$ | $y$ | $s_x$ | $s_y$ | $|s_x|$ | $|s_y|$ | $s$ |
| **2** | 0,00 | 0,00 |   |   |   |   |   |
| **3** | 0,20 | -0,03 | 0,20 | -0,03 | 0,20 | 0,03 | 0,21 |
| **4** | 0,27 | -0,05 | 0,07 | -0,02 | 0,07 | 0,02 | 0,07 |
| **5** | 0,16 | -0,46 | -0,12 | -0,41 | 0,12 | 0,41 | 0,43 |
| **...** | ... | ... | ... | ... | ... | ... | ... |

Натискання на клавішу F9 приводить до автоматичного перерахунку за новими даними (новими випадковими числами). Відповідно до цього змінюється вигляд траєкторії руху броунівської частинки. Отримувані у такий спосіб картинки можуть нагадувати сюжети з *рис. 1*.

**2.4. Статистичний аналіз результатів експерименту.** Кожен стовпець створеної таблиці містить випадкові числа, але не всі вони є зручними для аналізу. Зокрема, значення координат $x$ та $y$ лежать у широкому діапазоні з непередбачуваними границями. Зручнішими для аналізу є значення проекцій переміщення на осі координат $s_x$, або $s_y$, які потрапляють в інтервал [-1;+1]. Найбільш зручними виявляються модулі цих проекцій $|s_x|$ і $|s_y|$, що розташовані в іще більш вузькому інтервалі від 0 до +1.

То ж виконаємо нескладне статистичне дослідження випадкових чисел зі стовпця E. Насамперед виконаємо першу і обов'язкову процедуру статистичної обробки даних – їхнє *групування*, тобто розчленування на групи за певною ознакою. До першої групи включимо всі числа, менші за 0,1 (з інтервалу від 0 до 0,1); до другої – ті, значення яких знаходяться в інтервалі від 0,1 до 0,2, до третьої – числа з інтервалу 0,2 – 0,3 і т.д. – усього 10 груп.

Далі підрахуємо кількість чисел (елементів) у кожній із цих десяти груп. Для виконання такого завдання скористаємось функцією, яка в середовищі електронних таблиць у заданому діапазоні комірок підраховує кількість непустих комірок, вміст яких задовольняє заданій умові. Такою є функція

СЧЕТЕСЛИ(диапазон; "условие").

Тут діапазоном є адреси комірок, у яких розташовані випадкові числа, що їх ми маємо розбити на групи. Умова може бути задана, зокрема, за допомогою відношень «дорівнює» (=), «більше» (>), «менше» (<), «не більше» (<=), «не менше» (>=). Слід, однак, мати на увазі, що *умова не може* *бути складеною*, наприклад, не може бути ">5 і <10", вона має бути тільки простою.

Саме тому для підрахунку кількості елементів, які належать інтервалу від 0,1 до 0,2 виявляється неможливим створити, наприклад, конструкцію
СЧЕТЕСЛИ(АДРЕС1:АДРЕС2;">0,1;<0,2"),
а проблему вирішує конструкція
СЧЕТЕСЛИ(АДРЕС1:АДРЕС2;"<0,2") –
– СЧЕТЕСЛИ(АДРЕС1:АДРЕС2;"<0,1").

Останню формулу ілюструє рисунок

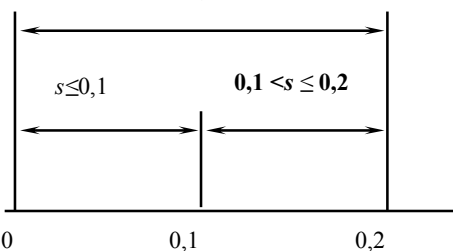

Створимо ще одну таблицю (*рис. 3*)

|   | I | J | K | L | M |
|---|---|---|---|---|---|
|   | Інтервали | | Середина інтервалу | Кількість в інтервалі | |
|   | від s ≥ | до s < | | абсолютна | відносна |
|   | 0,0 | 0,1 | 0,05 | 11 | 0,11 |
|   | 0,1 | 0,2 | 0,15 | 13 | 0,13 |
|   | ... | ... | ... | ... | ... |

*Рис. 3*

У стовпцях **I** та **J** показані границі інтервалів для кожної з десяти груп (дані у цих стовпцях уведені з клавіатури), стовпець **K** містить середини відповідних інтервалів, проте найбільш цікава й важлива інформація міститься у стовпцях **L** і **M**.

Вміст комірок у цих стовпцях наступний:

| комірки | формули / числа |
|---|---|
| K3 | =(I3+J3)/2 |
| L3 | =СЧЁТЕСЛИ(E3:E102;"<=0,1") |
| L4 | =СЧЁТЕСЛИ(E3:E102;"<0,2")– –СЧЁТЕСЛИ(E3:E102;"<=0,1") |
| M3 | =C3/1 |

Формули з комірок L4 та M3 копіювати в решту комірок відповідних стовпців з наступним редагуванням.

Експериментування тут зводиться до натискання на клавішу F9 (автоматичний перерахунок), внаслідок чого змінюється вміст усіх комірок обох таблиць.

Уміст стовпця **L**, нажаль, не дозволяє зробити ніяких висновків про яку-небудь певну закономірність у розподілі випадкових величин у групах. Той самий результат при бажанні можна побачити і на гістограмі, побудованій за даними стовпця **M**.

Зауважимо, що математична статистика вивчає численні сукупності елементів, і чим більше елементів містить сукупність, тим більш надійними й адекватними виявляються результати статистичного дослідження. Саме тому кількість рядків (елементів) у всіх стовпцях від **A** до **G** попередньої таблиці 2 доцільно збільшити, як покажуть досліди, від 100 до хоч би 5000. Як завжди, здійснимо це копіюванням формул останнього рядка з номером 102 до рядка з номером 5002. Після кількох натискань на F9 спостерігаємо таке чи подібне:

|   | I | J | K | L | M |
|---|---|---|---|---|---|
|   | від s ≥ | до s< | | абсолютна | відносна |
|   | 0,00 | 0,10 | 0,05 | 535 | 0,107 |
|   | 0,10 | 0,20 | 0,15 | 497 | 0,099 |
|   | 0,20 | 0,30 | 0,25 | 480 | 0,096 |
|   | 0,30 | 0,40 | 0,35 | 519 | 0,104 |
|   | 0,40 | 0,50 | 0,45 | 480 | 0,096 |
|   | 0,50 | 0,60 | 0,55 | 519 | 0,104 |
|   | 0,60 | 0,70 | 0,65 | 522 | 0,104 |
|   | 0,70 | 0,80 | 0,75 | 470 | 0,094 |
|   | 0,80 | 0,90 | 0,85 | 459 | 0,092 |
|   | 0,90 | 1,00 | 0,95 | 519 | 0,104 |

Тепер остання таблиця має поновлений вигляд, і нарешті ця таблиця разом з відповідною гістограмою (*рис. 4*)

дозволяє встановити, що розподіл випадкових чисел за визначеними десятьма групами є майже рівномірним. Таким самим є розподіл випадкових чисел у всіх решта стовпцях попередньої таблиці (координат, проекцій переміщення, модулів цих проекцій тощо).

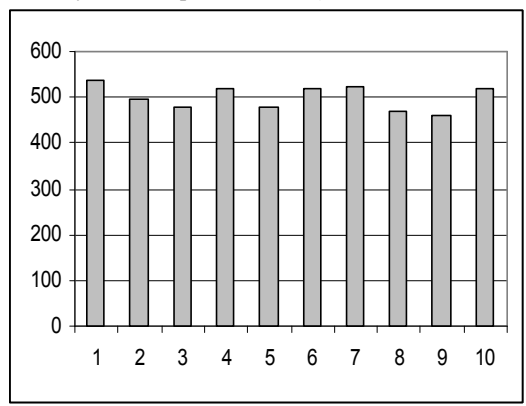

*Рис. 4*

Випадкові числа, які продукує комп'ютер, є рівномірно розподіленими: будь-якому значенню випадкової величини відповідає одна й та сама ймовірність появи.

У природі зазвичай усяка мінливість розподіляється нерівномірно, і, скоріш за все, не існує фізичних процесів, які б могли бути описані за допомогою рівномірного розподілу.

**2.5. Приклад природного розподілу.** В кабінеті шкільного лікаря зберігаються медичні карти кожного школяра, де міститься чимало медичних і фізіологічних покажчиків. Серед них розглянемо один – зріст. Візьмемо навмання групу учнів деякого класу, і зріст (у сантиметрах) кожного з 30 школярів упишемо до таблиці, але не за абеткою, а заздалегідь впорядкувавши.

| 143 | 150 | 155 | 158 | 163 |
| 144 | 151 | 155 | 160 | 164 |
| 146 | 152 | 156 | 161 | 166 |
| 147 | 153 | 156 | 161 | 168 |
| 148 | 153 | 156 | 161 | 169 |
| 150 | 155 | 157 | 162 | 171 |

*Рис. 5*

Виконаємо поділ отриманих даних на групи шириною 5 см: перша від 140 до 144 см, друга від 145 до 149 см і т.д.

*Примітка.* Задавати інтервали рекомендують так, щоб їхня кількість $k$ була не меншою за 6 і не більшою 20.

Тепер заповнимо наступну таблицю 6:

|   | A | B | C | D | E |
|---|---|---|---|---|---|
|   | Інтервали | | Середина інтервалу | Кількість в інтервалі | |
|   | від s >= | до s < | | абсолютна | відносна |
|   | 140 | 144 | 142 | 2 | 0,067 |
|   | 145 | 149 | 147 | 3 | 0,100 |
|   | 150 | 154 | 152 | 6 | 0,200 |
|   | 155 | 159 | 157 | 8 | 0,267 |
|   | 160 | 164 | 162 | 7 | 0,233 |
|   | 165 | 169 | 167 | 3 | 0,100 |
|   | 170 | 174 | 172 | 1 | 0,033 |

Комірки у стовпцях **A, B, C** таблиці заповнюються з клавіатури згідно з даними таблиці за *рис. 5*. Стовпець **D** можна заповнювати або за формулами стовпця **L** таблиці з *рис. 3*, або простим підрахунком за таблицею на *рис. 5* завдяки малій кількості елементів у ній. Формули у комірках стовпця E не повинні викликати утруднень.

Будуючи гістограму за даними стовпця **D** *таблиці 6*, отримуємо наступний розподіл росту за сьома визначеними групами (*рис. 6а*). Цей природний розподіл докорінно відрізняється від рівномірного, він є близьким до так званого *нормального* розподілу або розподілу Гауса. Він є так само ідеалізованим, як і розглянутий перед цим рівномірний, функція цього розподілу має вигляд симетричної дзвоноподібної кривої, що асимптотично наближається до вісі абсцис (*рис. 6б*).

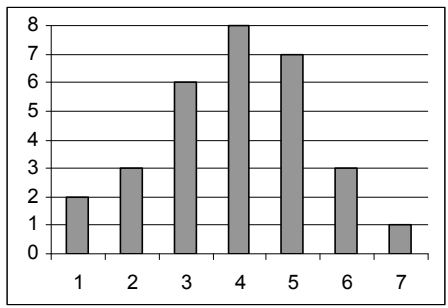

*Рис. 6а*

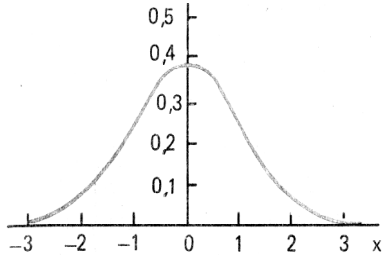

*Рис. 6б*

**3. Моделювання за допомогою нормально розподілених випадкових чисел.**

Електронні таблиці дозволяють генерувати не тільки рівномірно розподілені випадкові числа, але й випадкові числа за деякими іншими найчастіше вживаними розподілами. Як отримати такий розподіл в середовищі електронних таблиць, можна дізнатися в [8, 231].

**3.1. Картини броунівського руху з нормальним розподілом окремих випадкових переміщень**. Отже, створимо два стовпця нормально розподілених випадкових чисел по сто чисел у кожному. Ці числа моделюватимуть переміщення $\Delta x$ і $\Delta y$ броунівської частинки. Координати $x_i$ та $y_i$ частинки на будь-якому проміжку часу з номером $i$, як і раніше, знайдемо так:

$$x_i = x_{i-1} + \Delta x; \quad y_i = y_{i-1} + \Delta y.$$

Початковим координатам знову надамо нульових значень: $x_0 = 0$; $y_0 = 0$.

Наведемо можливий варіант заповнення такої таблиці і за даними стовпців $x$ і $y$ побудуємо траєкторію руху частинки – графік $y = y(x)$.

|   | A | B | C | D |
|---|---|---|---|---|
| 1 | Δx | Δy | x | y |
| 2 | 0 | 0 | 0 | 0 |
| 3 | 1,601 | -1,457 | 1,601 | -1,456 |
| 4 | 0,276 | -0,425 | 1,876 | -1,880 |
| 5 | -0,296 | -0,367 | 1,580 | -2,247 |
| ... | ... | ... | ... | ... |

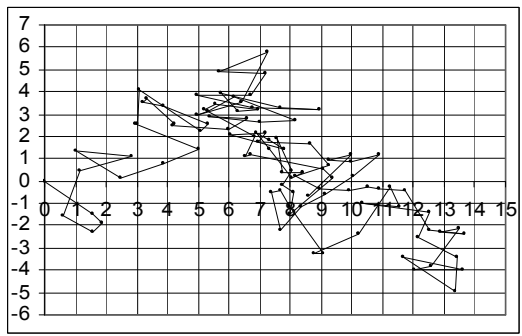

*Рис. 7*

Повторюючи процедуру отримання нормально розподілених випадкових чисел (переміщень $\Delta x$ і $\Delta y$), у нових таблицях можна так само добудувати стовпці C і D для поточних координат $x$ та $y$ і вивести на екран нові траєкторії.

Картини на нових рисунках знов нагадують сюжети з *рис. 1*. Візуальне порівняння цих рисунків з *рис. 2* не виявляє суттєвої різниці між ними. Отже для *ілюстрування* броунівського руху генерування випадкових чисел за рівномірним або нормальним розподілом не є суттєвим. Але оскільки функція для рівномірного розподілу СЛЧИС( ) реалізується простіше і здатна до автоматичного перерахунку всього лише одним натисканням на клавішу F9, то їй зазвичай віддають перевагу.

*Зауваження*. Для того, щоб отримати картинки, подібні до тих, що наведені на *рисунках 2* або *7*, треба виконувати декілька експериментів і, можливо, почекати, поки не з'явиться придатний (гарний) рисунок.

*Вправи.*

1. У чому полягає ідея методу Монте-Карло?
2. Чи повинні збігатися значення змінних у поданих тут таблицях з відповідними даними у таблицях, створених вами?
3. Для чого використовують рівномірно розподілені випадкові числа?
4. Коли використовують нормально розподілені випадкові числа?
5. Запропонуйте функцію для отримання в електронних таблицях однозначних цілих випадкових чисел в інтервалі [–9; 9] за допомогою функції СЛЧИС( ).
6. Виконайте статистичне дослідження даних зі стовпця G таблиці 6.

**Висновки**:

1. Випадкові числа, які продукує комп'ютер, зокрема в середовищі електронних таблиць за допомогою функції СЛЧИС( ), є рівномірно розподіленими, тобто будь-якому значенню випадкової величини відповідає одна й та сама ймовірність. На практиці такий розподіл використовують при комп'ютерному моделюванні складних систем у якості основи при побудові стохастичних моделей.

2. Побудована нами модель броунівського руху із застосуванням рівномірно розподілених випадкових чисел виявилася вдалою тільки на перший погляд, тільки на якісному рівні. Адже з опрацювання результатів фізичних спостережень та з дослідів добре відомо, що особливості такого руху характеризуються не рівномірним, а нормальним законом розподілу. В електронних таблицях також є засоби генерування нормально розподілених випадкових чисел.

3. В математичній статистиці, окрім розглянутих тут рівномірного і нормального розподілів, відомі й інші розподіли, не менш важливі.

**Перспективи подальших досліджень:** розробка методичних основ навчання імітаційного комп'ютерного моделювання у середовищі електронних таблиць.

This article is devoted to methods of construction and study of stochastic models based on Monte Carlo method. A model of Brownian motion, the construction and processing which brings to a world of random numbers and mathematical statistics, promotes understanding of the probability distribution, in particular illustrates two common distributions: uniform and normal.

**Key words**: computer simulation, Monte Carlo method, random numbers, uniform distribution, spreadsheets.